\documentclass[aps,preprint,floatfix,nofootinbib]{revtex4}
\usepackage{graphicx}
\usepackage{epsf}

\begin{document}  

\preprint{MADPH--04--1375}
\preprint{hep-ph/0405055}

\vskip 1cm
\title{Hadron Collider Signatures\\  
for New Interactions of Top and Bottom Quarks}

\author{T. Han$^a$,\ G. Valencia$^b$,\ Yili Wang$^b$}

\email{than@pheno.physics.wisc.edu}
\email[]{valencia@iastate.edu}
\email[]{yiliwa@iastate.edu}

\affiliation{$^a$Department of Physics, University of Wisconsin, Madison, 
WI  53706, and \\
Institute of Theoretical Physics, Academia Sinica, Beijing 100080, China\\
$^b$Department of Physics, Iowa State University, Ames, IA 50011}

\date{\today}

\vskip 1cm
\begin{abstract} 
  
One of the main goals for hadron colliders is the study of the 
properties of the third generation quarks. We study the signatures 
for new TeV resonances that couple to top or bottom quarks both 
at the Tevatron Run II and at the LHC. We find that in the simplest 
production processes of Drell-Yan type at the Tevatron,
the signals are overwhelmed by QCD backgrounds.
We also find that it is possible to study these resonances when they 
are produced in association with a pair of heavy quarks or in 
association with a single top at the LHC.
In particular, with an integrated luminosity of 300 fb$^{-1}$
at the LHC, 
it is possible to probe resonance masses up to around 2~TeV. 

\end{abstract}


\maketitle

\section{Introduction}  

A major goal for the Fermilab Tevatron and the CERN LHC is the detailed study 
of the properties of the top quark. In particular they should 
establish whether the third 
family behaves like the first two, or whether it is subject to new 
interactions. The interactions of the third family have been studied 
indirectly in many low energy processes such as rare kaon and B meson decays  
with no firm evidence for physics beyond the standard model. There are, 
however, certain inconsistencies \cite{chano} associated with the 
forward-backward asymmetry $A^b_{FB}$ measured at LEP 
that hint at a potential problem. In any case it is highly desirable to pursue 
a direct study of the couplings of the top quark in colliders. 
With this in mind, and motivated by the possibility 
that the top quark plays a special role in the breaking of electroweak 
symmetry \cite{topcolor},
we have previously studied the signals for a new resonance 
in the process $W_LW_L \rightarrow t \bar{t}$ \cite{han}. 

A different type of signal occurs if the new resonances couple strongly 
to the third generation of quarks but not necessarily to the $W$ and $Z$ 
gauge bosons. 
In this paper we study the signatures for this type of new physics 
at the Tevatron and at the LHC. We first discuss the cases of a vector  
and a scalar resonance (generically denoted by $R$) 
produced in the $s$-channel processes  
$q\bar q \rightarrow R\to  b\bar{b} {\rm ~or~} t \bar{t}$. In these processes 
the QCD 
backgrounds are large and we apply known techniques to reduce these 
backgrounds. 
We then consider the production of the new resonances 
in association with a $b\bar{b}$ or 
$t\bar{t}$  pair through processes such as 
$gg \rightarrow R t \bar{t} \rightarrow b \bar{b}t \bar{t}$ or 
$t \bar{t}t \bar{t}$. 
These processes are higher order corrections to the $s$-channel production 
and therefore have a significantly smaller cross-section. However, 
their unique topology permits a much better control of the QCD background 
and we find that they yield potentially observable signals.

In our study of the process $W_LW_L \rightarrow t \bar{t}$ \cite{han} 
we used model-independent (but non-renormalizable) parameterizations for the 
couplings of the new resonances to the $b$ and $t$ quarks. We found that if 
these resonances were responsible for electroweak symmetry breaking they were 
typically very broad when they were as heavy as a few TeV. In order to 
parametrize new resonances that are heavy and narrow we argued in 
Ref.~\cite{han} that we had to consider models 
with more than one new resonance at a time. In this paper we adopt this 
scenario and do not require that our resonances be responsible for electroweak 
symmetry breaking. We say nothing of their couplings to $W$ and $Z$ 
gauge bosons but simply assume that they provide negligible contributions 
to their widths. A prototype for a vector resonance with such behavior: 
no coupling to the $W$ and $Z$ gauge bosons; very weak coupling to the 
first two generations of fermions; and large 
coupling to $b$ and $t$ quarks is provided by the 
$Z^\prime$ from Ref.~\cite{zprime}. For definiteness we will 
have this particle in mind as our vector resonance. 
We can regard this as a fairly general parametrization of a new vector
resonance with arbitrary couplings to $b$ and $t$ if we abandon 
renormalizability and treat the resonance couplings as an effective 
theory, in the spirit of Ref.~\cite{han}. For the 
case of a scalar resonance we use the simple parameterization of 
the $Sb\bar{b}$ and $St\bar{t}$ couplings that we used in 
Ref.~\cite{han} but we assume a negligible 
width $\Gamma(S \rightarrow WW)$.  
The possibility of non-standard couplings for the top quark has 
been discussed extensively in the literature in the context of 
anomalous couplings \cite{topac}. 
Some generic models of vector resonances coupled to the top quark
strongly are studied in Ref.~\cite{hillparke}.
By adding new resonances as explicit 
degrees of freedom to the effective theory, one is able to study 
potential non-standard couplings in a larger energy domain. 

\section{Model for New strongly interacting Resonances}\label{sec:model} 

\subsection{Vector Resonance}

We begin by discussing our parametrization for the new vector resonance. 
Effective interactions between the SM gauge bosons and fermions and new 
vector resonances have been described in the literature~\cite{bess}. 
We are interested in new interactions 
of $b$ and $t$ quarks to new heavy resonances that are sufficiently 
narrow to be described by a Breit-Wigner shape. One way to accomplish this 
within the effective Lagrangian framework is to have more than one resonance 
as we discussed in Ref.~\cite{han}. Here we assume that this is 
the case, and that the resonance under study has negligible couplings 
to electroweak gauge bosons. An alternative way to obtain such a resonance 
is to consider extended gauge sectors. Our goal in this paper is to 
investigate the extent to which hadron colliders are sensitive to new 
strong interactions of the top quark regardless of the origin of the 
new interactions. We thus proceed with the following effective Lagrangian 
coupling a spin one field to the top and bottom quarks,
\begin{equation}
{\cal L} = - \Psi \gamma^\mu (g_V + g_A \gamma_5)\tau_i \Psi V^i_\mu.
\label{simplevp}
\end{equation}
Vector resonances introduced in this manner occur in the BESS model, 
for example \cite{bess}. They 
have universal couplings to fermions that arise from mixing between 
the new vectors and the $W$ and $Z$. In addition there can be non-universal 
direct couplings. We are interested only in the latter in the form of 
large new couplings to the $b$ and $t$ quarks because universal 
couplings are severely constrained by low energy observables. In fact, the 
most stringent constraints on the couplings in Eq.~(\ref{simplevp})  were 
found in Ref.~\cite{han} to arise from mixing between the new $V$ 
and the $W$ and $Z$. A direct study of these couplings at high energy will 
therefore be most relevant for the case of negligible mixing and we 
concentrate on this case.

In hadron colliders, however, light quark annihilation represents a 
significant production source for new vector resonances even if they 
couple predominantly to $b$ and $t$. This is an unfortunate complication 
because it forces us to commit to a specific model where the relative 
couplings between the new vectors and the heavy and light quarks are known. 
To keep our study as model-independent as possible we will illustrate our 
results for parameters that make the contributions of the light quark 
annihilation mechanism to resonance production small. We will also compare 
with a scalar resonance in which we assume no couplings to light fermions 
exist. For definiteness we will use the $Z^\prime$ model of 
Ref.~\cite{zprime} in the limit of no $V-Z$ mixing. There exist 
very tight constraints on the flavor changing neutral currents that appear 
in this model \cite{zprime} and therefore, our starting point in this 
paper will be the flavor diagonal interaction in the quark mass eigenstate 
basis, 
\begin{eqnarray} 
{\cal L} &=& 
{g\over 2}\tan\theta_W \tan\theta_R  
({1\over 3} \bar q_L \gamma^\mu q_L+ {4\over 3} \bar u_{Ri} \gamma^\mu u_{Ri}
-{2\over 3} \bar d_{Ri}\gamma^\mu d_{Ri}) V_\mu \nonumber\\
&-& {g\over 2}\tan\theta_W  (\tan\theta_R + \cot\theta_R) (
\bar t_{R} \gamma^\mu t_{R} - 
\bar b_{R} \gamma^\mu b_{R}) V_\mu . 
\label{neucoupl}
\end{eqnarray}
In this expression $g$ is the standard model $SU(2)_L$ gauge coupling, 
$\theta_W$ is the usual weak mixing angle and $\theta_R$ is a new 
parameter. Also $q_L$ is summed over  the SM quarks, 
and repeated indices are summed over the three generations. 
With large $\cot\theta_R$, this model provides a specific 
example of a new vector resonance with 
couplings to $b$ and $t$ that are significantly enhanced with respect to 
couplings to the light fermions. In the limit of large $\cot\theta_R$ 
these couplings are purely right-handed, with 
\begin{equation}
g_A = g_V = {g \over 4}\tan\theta_W\cot\theta_R 
\end{equation} 
We have checked numerically that the signals discussed in this 
paper would be very similar if we had left handed couplings instead. 

The resonance width into $b\bar{b}$ or $t\bar{t}$ pairs 
is \footnote{In Refs.~\cite{han} we used an equation 
for this width that has a typo. Numerically it does not affect our 
conclusions in those papers because it only affects terms that are 
suppressed by $m_t^2/M_R^2$ and we considered large resonance masses.} 
\begin{eqnarray}
\Gamma(V \rightarrow  f \bar{f}) &=& {M_V \over 2 \pi}
g_V^2
\left(1-4{m_f^2\over M_V^2}\right)^{1/2} 
\left(1-{m_f^2\over M_V^2} \right)\\
&\approx&  63\ 
\left( { g_V\over 0.63 } \right)^2
\left( {M_V \over 1000\ {\rm GeV}} \right) {\rm~GeV},
\quad {\rm for}\ M_V\gg m_f .
\end{eqnarray}
Requiring the new interaction to remain perturbative 
leads to the theoretical constraint $\cot\theta_R < 20$, 
equivalently $g_V < 1.8$. Partial 
wave unitarity requirements do not improve this bound. 
For illustration purposes, throughout the paper, 
we will present numerical results with 
\begin{equation}
g_V = 0.63 {\rm ~\Longleftrightarrow~ }\cot\theta_R =7. 
\label{cotr}
\end{equation}
With this choice the couplings of $V$ to $b$ and $t$ are 
about 50 times larger than its couplings to the light quarks. At the 
same time the resonance remains narrow, with $\Gamma_V/M_V \sim 0.12$. 
Notice that the value  $g_V \sim 0.63$ is about four 
times larger than the largest value we considered in Ref.~\cite{han}.  
In that case we were constrained by low energy bounds on $V-W$ mixing, 
whereas here we consider the case where that mixing is independent 
from the coupling $g_V$ and effectively remove the constraints. 

The relative branching fraction for the decays of the $V$ into  $b\bar{b}$ and 
$t\bar{t}$ is governed simply by kinematics, 
\begin{equation}
\frac{\Gamma(V \rightarrow t\bar{t})}{\Gamma(V \rightarrow b\bar{b})}  = 
\left (\frac{M_V^2 - 4 m_t^2}{M_V^2 - 4 m_b^2}\right )^{1/2} 
\left(\frac{M_V^2 -  m_t^2}{M_V^2 -  m_b^2 }\right).
\label{br-vec}
\end{equation}
For small $M_V$, $\Gamma(V \rightarrow b\bar{b})$ is much larger 
than $\Gamma(V \rightarrow t\bar{t})$. As $M_V$ nears $500$~GeV, 
$\Gamma(V \rightarrow t\bar{t})$ is only about $62\%$ of 
$\Gamma(V \rightarrow b\bar{b})$ and increases to $90\%$ 
when $M_V$ is near 1~TeV.

\subsection{Scalar Resonance}

We next consider the effective interaction between the 
third generation quarks and a new scalar resonance. In this case we will use 
a very simple (non-renormalizable) parametrization for the new interactions,  
and assume that the couplings of the scalar to the light fermions are 
completely negligible. We write 
\begin{equation}
{\cal L}= - {m_t\over v} S \left( \kappa_b \bar b b + \kappa_t \bar t t\right).
\label{lstt}
\end{equation}
This form allows us to parameterize simultaneously the cases 
where either the $b$-quark or the $t$-quark or both have enhanced 
couplings to the new scalar resonance. Examples where the $b$ quark coupling 
(rather than the $t$ quark coupling) to a scalar is enhanced occur 
frequently in multi-Higgs models with large $\tan\beta$. 
We will assume in this study that the scalar width is dominated by its 
decay into $b$ and $t$ pairs and that it receives a negligible contribution 
from decay into $W$ and $Z$ pairs. This corresponds to taking 
$g_S \sim 0$ in the models discussed 
in Ref.~\cite{han} and therefore implies the 
existence of additional new resonances to restore unitarity in 
$WW$ scattering amplitudes. The new couplings $\kappa_{b,t}$ are 
related to the width of the scalar into quark pairs,
\begin{eqnarray}
\Gamma_{Sf\bar{f}} &=& 
{3\kappa_f^2\over 8\pi}{m_t^2 M_S \over v^2}
\left(1-{4m_f^2\over M_S^2}\right)^{3\over 2}\; \\
&\approx&  60\  \kappa_f^2
 \left( {m_t\over 175\ {\rm GeV}} \right)^2
\left( {M_S \over 1000\ {\rm GeV}} \right){\rm GeV}, 
\quad {\rm for}\ M_S\gg m_f .
\label{stt}
\end{eqnarray}

As argued in Ref.~\cite{han} there 
are few constraints on these couplings from low energy observables. 
The tightest constraint is obtained by requiring perturbative 
unitarity~\cite{han,unit} in the scattering amplitude 
$b\bar{b} \rightarrow b \bar{b}$ 
(or in $t\bar{t} \rightarrow t \bar{t}$) 
through an exchange of the new scalar. This leads to 
\begin{equation}
\kappa_{b,t} \leq 3.
\end{equation}

The relative branching fraction for the new scalar resonance decay to 
$b\bar{b}$ and $t\bar{t}$ 
pairs is a free parameter depending on the square of the ratio 
$\kappa_b/\kappa_t$. 
For our numerical  illustrations, we will use
\begin{equation}
\kappa_b = \kappa_t = 1.
\label{kappa}
\end{equation}
This appears to be an unusual choice because in many models with 
couplings such as Eq.~(\ref{lstt}) one does not have simultaneously 
large $\kappa_b$ and $\kappa_t$. However, as we will see, different 
processes that we consider single out one of the couplings. The value 
we have chosen for these couplings results in an interaction that is 
weaker than it was in the vector case. A more detailed study would 
determine the sensitivity to the couplings in both scalar and vector 
case. For $M_S \sim 500$~GeV, $\Gamma(S \rightarrow t\bar{t})$ 
is only about $36\%$ of $\Gamma(S \rightarrow b\bar{b})$, while for 
$M_S \sim 1$~TeV, it increases to $82\%$. Our choice of parameters 
will allow us to illustrate two slightly different cases, the scalar 
will be narrower than the vector and it will not receive any contribution 
from light quark annihilation.

\section{Heavy $q\bar{q}$ pair production at the Tevatron}

\begin{figure}[tb]
\begin{center}
\includegraphics[width=14cm]{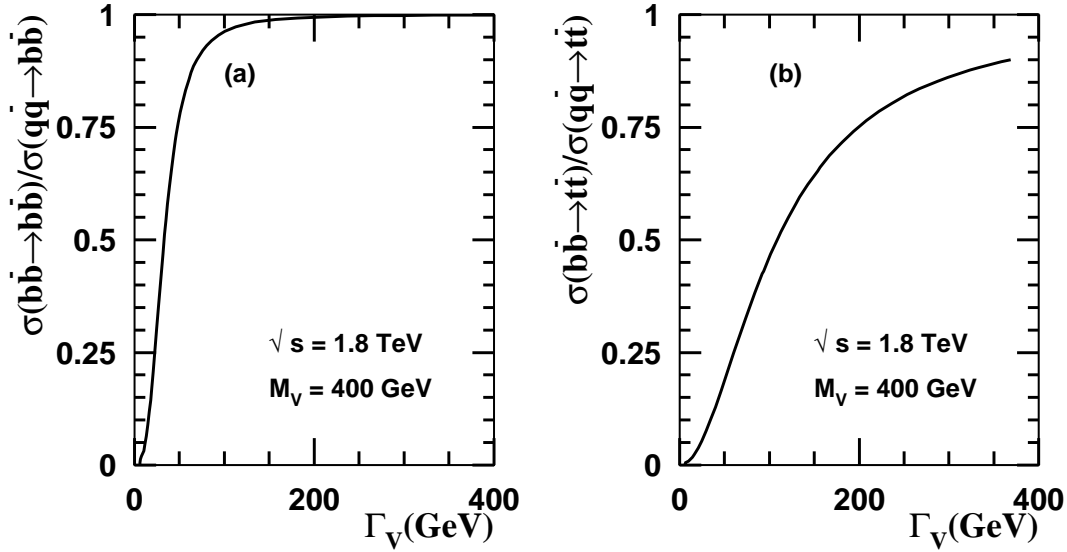}
\end{center}
\caption{Relative contributions (a) to
$p\bar{p} \rightarrow b \bar{b}X$ and (b) to  $t \bar{t}X$
from $b\bar{b} \rightarrow b\bar{b}\ (t\bar t)$ and 
$q\bar{q} \rightarrow b\bar{b}\ (t\bar t)$ 
processes at the Tevatron for $M_V =400$~GeV as a function of $\Gamma_V$.}
\label{fig:bftev}
\end{figure}

The processes $p\bar{p} \rightarrow b \bar{b}X,\ t\bar tX$ receive enhanced 
contributions
from the parton level processes $b\bar{b} \rightarrow b\bar{b},\  t\bar t$ 
involving the $s$-channel 
exchange of the new resonance. The vector resonance also 
receives corrections of electroweak strength from the parton level processes 
$q\bar{q} \rightarrow b \bar{b},\ t\bar t$ for a light quark $q$. 
These contributions can actually be dominant in some regions of parameter space 
because the light quark content of the proton is much larger than its 
$b$ content. To the extent that this is possible, we will only consider 
cases where the $b\bar{b}$ annihilation mechanism dominates to keep our 
conclusions as model-independent as we can.

To this effect we compute the relative contributions of both mechanisms 
to the processes  $p\bar{p} \rightarrow b\bar{b}X $ and 
$p\bar{p} \rightarrow t\bar{t}X$. We present the results in 
Fig.~\ref{fig:bftev}(a) and Fig.~\ref{fig:bftev}(b) respectively.
We select a resonance mass $M_V =400$~GeV which is above threshold for 
$t\bar{t}$ production and about as large as can be probed by the 
Tevatron. We present our results as a function of the resonance 
width $\Gamma_V$. 
Increasing the width corresponds to increasing the coupling 
to the top and bottom quarks while reducing the coupling to the light 
quarks. For this reason the fraction of signal events that originates 
in $b\bar b$ annihilation increases as a function of $\Gamma_V$. 
We find that in $b\bar{b}$ production, at $\Gamma_V \ge 45$~GeV, more than 
$80\%$ of the signal is coming from $b\bar b$ annihilation. Since our 
motivation is to study the couplings to the third generation, we choose  
$\cot\theta_R = 7$ or $g_V =0.63$ as in Eq.~(\ref{cotr}),
 corresponding to $\Gamma_V \sim 47$~GeV  for our Tevatron studies. 

For $t\bar t$ production on the other hand, light quark annihilation 
represents a larger fraction of signal events than for $b\bar{b}$ production.  
Figure~\ref{fig:bftev}(b) indicates that $\Gamma_V \ge 100$~GeV 
would be necessary for $b \bar{b}$ annihilation to produce 50\% of the 
signal,  and about $\Gamma_V \ge 200$~GeV for it to dominate. These values, 
however, correspond to unacceptably large couplings. We thus keep 
$\Gamma_V \sim 47$~GeV. In this case only 17\% of the signal events 
are produced through the couplings that we want to study. We will be 
able to improve this situation with the higher energy available at 
the LHC.

For the scalar resonance the signal arises exclusively from 
$b\bar b$ annihilation for both $b\bar b$ and $t\bar t$ production 
because we are ignoring the couplings to the light quarks. 
It is evident that $b\bar b$ production will only be sensitive to 
$\kappa_b$, whereas $t\bar t$ production will be sensitive to the 
product $\kappa_b \kappa_t$. As we mentioned in the Introduction, 
many scalars found in models commonly discussed have enhanced 
couplings to $b$ or $t$ but not to both at the same time. For those 
cases we would not expect a signal in $t\bar t$ production.
For illustration we choose $\kappa_b = \kappa_t = 1$, 
as in Eq.~(\ref{kappa}),  well below the 
unitarity constraint. For a 400~GeV scalar mass this then corresponds 
to $\Gamma_S = 27$~GeV. With these parameters, our scalar resonance 
is roughly half as narrow as our vector resonance and we expect 
relatively fewer signal events. 

\begin{figure}[tb]
\includegraphics[width=14cm]{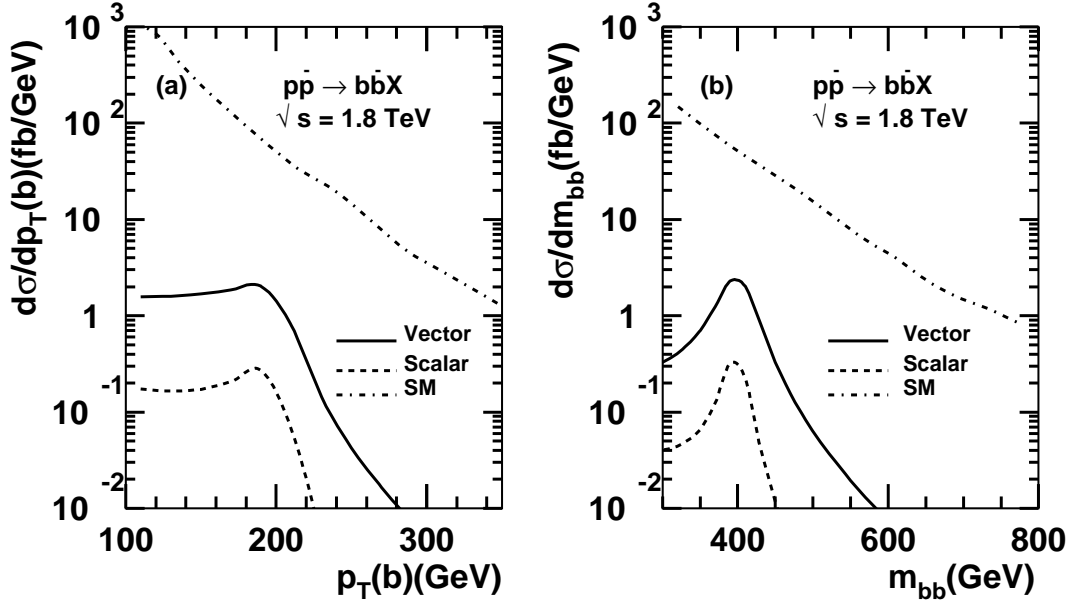}
\caption{(a) $p_T(b)$ and (b) $m^{}_{b\bar{b}}$ distributions for the 
signal and background at the Tevatron. 
The solid curves correspond to the  signal for a vector resonance 
with $M_V=400$~GeV and $\Gamma_V = 47$~GeV. 
The dashed curves correspond to the signal for a scalar resonance
with $M_S=400$~GeV and  $\Gamma_S = 27$~GeV. 
The dot-dashed curves are standard model background. }
\label{fig:tevpt}
\end{figure}

We first consider $b\bar{b}$ production 
and demand both $b$'s to be tagged. We assume a 
combined efficiency of $50\%$ (or about $70\%$ for each 
tagged $b$) \cite{rate}. We compute the standard model background with 
the aid of MADGRAPH \cite{Stelzer:1994ta}. 
We include both the physical background consisting of 
QCD produced $b\bar{b}$ pairs, as well as a fake background 
that results when final state light quarks mimic a $b\bar{b}$ pair. 
The rate at which this occurs is assumed to be $0.5\%$ \cite{rate}. 
The main production mechanism for the background $b\bar{b}$ pairs 
at the Tevatron is the QCD process via gluon fusion. The background 
$b\bar{b}$ pairs have mostly low or intermediate transverse 
momentum and we adopt a $p_T$ cut to suppress them. We also 
adopt a rapidity cut that mimics the typical coverage of D0 and 
CDF. Our basic cuts for Tevatron processes will thus be
\begin{eqnarray}
p_T(b) > 100 {\rm ~GeV},  \quad  |y_b| & <  2.
\label{tev-cuts}
\end{eqnarray}

We show the transverse momentum $p_T(b)$ and the invariant mass $m_{b\bar{b}}$ 
distributions in Figs.~\ref{fig:tevpt}(a) and (b) respectively. 
Unfortunately the background is several orders of magnitude 
larger than the signal and we were not able to find 
a way to reduce it significantly while preserving the signal.
To compute the statistical sensitivity of this process we 
optimize the signal/background ratio with the cut that 
discards events more than two widths away from the resonance mass,
\begin{eqnarray}
M_R - 2\Gamma_{R}  <  m_{b\bar{b}} <  M_R + 2\Gamma_{R}.
\label{tev-cuts1}
\end{eqnarray}

\begin{figure}[t]
\begin{center}
\includegraphics[width=14cm]{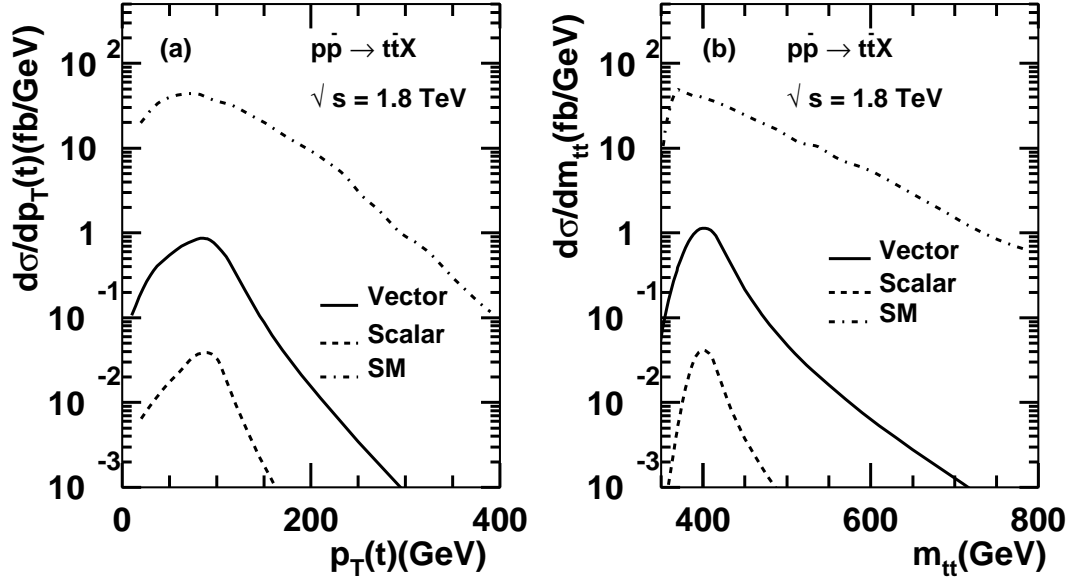}
\end{center}
\caption{(a) $p_T(t)$ and (b) $m^{}_{t\bar{t}}$ distributions for the 
signal and background at the Tevatron. 
The solid curves correspond to the  signal for a vector resonance 
with $M_V=400$~GeV and $\Gamma_V = 47$~GeV. 
The dashed curves correspond to the signal for a scalar resonance
with $M_S=400$~GeV and  $\Gamma_S = 27$~GeV. 
The dot-dashed curves are standard model background. }
\label{fig:tevpt-tt}
\end{figure}

We now turn our attention to  $t\bar{t}$ production. 
As was the case with $b\bar{b}$ production, the main background is 
QCD production of $t\bar{t}$ pairs. In this case the background is also 
two orders of magnitude larger than the signal. 
In Fig.~\ref{fig:tevpt-tt}, we show the top transverse momentum $p_T(t)$ 
and invariant mass $m_{t\bar{t}}$ distributions for both signal and 
background. For the signal we use the same model-parameters 
we used in $b\bar{b}$ production. We also implement the kinematical cut 
\begin{eqnarray}
|y_t| & <  2.
\label{tev-cutst}
\end{eqnarray}
Once again we optimize the sensitivity to 
new physics by selecting the  $m_{t\bar{t}}$ invariant mass region around the 
resonance as per Eq.~(\ref{tev-cuts1}).

\begin{figure}[tb]
\begin{center}
\includegraphics[width=14cm]{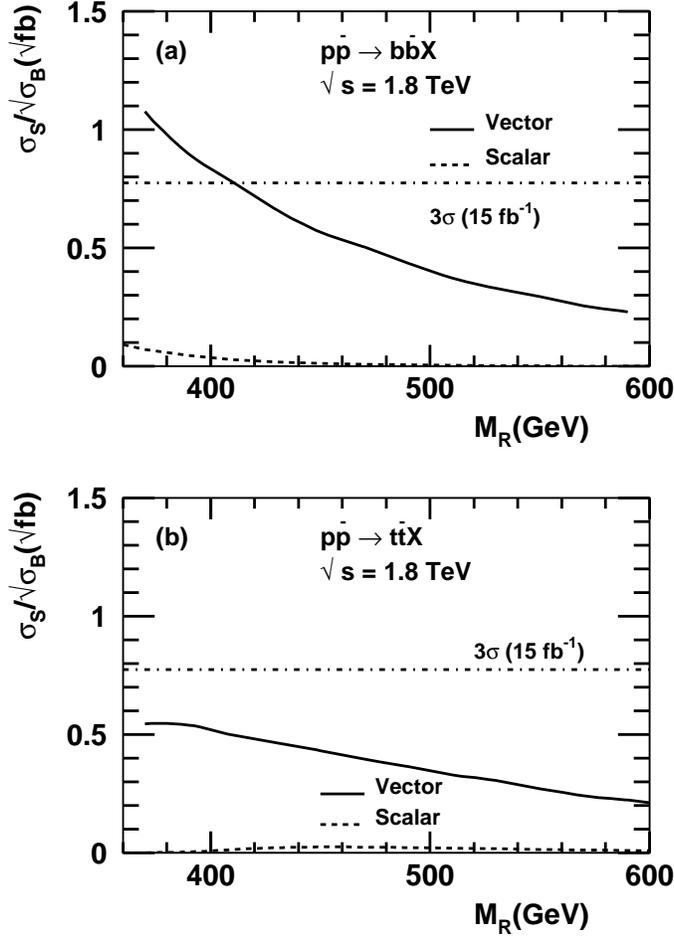}
\end{center}
\caption{ 
Statistical sensitivity to a new resonance as a function of the 
resonance mass for (a) $b\bar b$ production, and (b) $t\bar t$ production. 
The solid curve is for a vector resonance and the dotted curve for a scalar 
resonance. The dot-dashed line indicates the $3\sigma$ sensitivity level. 
We assume a Tevatron Run II  integrated luminosity of 15 fb$^{-1}$.}
\label{fig:tevreach}
\end{figure}

We estimate the statistical sensitivity to the signal by simply dividing 
the number of signal events by the square root of the number 
of the background events: 
\begin{eqnarray}
{\sigma_S\over \sqrt{\sigma_B} }\sqrt L,
\label{sig}
\end{eqnarray}
where $L$ is the total integrated luminosity. 
We show this statistical sensitivity  for the Tevatron Run II 
as a function of the resonance mass in Fig.~\ref{fig:tevreach}. 
To obtain our numbers we are using semileptonic channels for signal 
identification. One of the $t$ quarks decays leptonically into an 
electron or a muon and the other one decays hadronically. With 
50\% $b\bar b$-tagging efficiency we end up using a combined event 
efficiency of about $16\%$. 
The dot-dashed line in the figure indicates the $3\sigma$ signal sensitivity 
assuming a total integrated luminosity of 15 fb$^{-1}$. 
This figure indicates that it might be possible to observe a $3\sigma$ 
signal for a resonance lighter than about $M_R<400$ GeV in 
$b\bar b$ production. However, given the very low signal to background 
ratio more realistic studies at the detector level would be needed to 
conclude that this is observable at the Tevatron. We also conclude that 
the Tevatron is not sensitive to this type of new physics through the 
$t\bar{t}$ channel. In all cases we used couplings given by 
Eqs.~(\ref{cotr})~and~(\ref{kappa}) and we do not expect the conclusions
to change if we choose the model-parameters differently. 
The better sensitivity to a vector 
resonance is in large part due to its additional production mechanism 
through light quark annihilation.  As mentioned in the introduction these 
couplings to light quarks are very model-dependent.

\section{Signals at the LHC}

\begin{figure}[tb]
\begin{center}
\includegraphics[width=14cm]{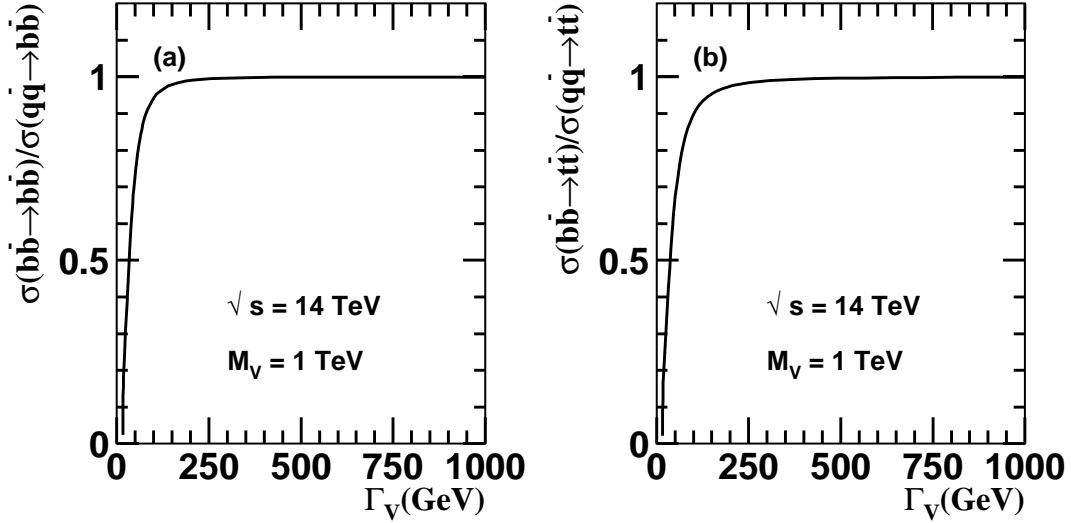}
\end{center}
\caption{
Relative contribution (a) to
$p{p} \rightarrow b \bar{b}X$ and (b) to  $t \bar{t}X$
from $b\bar{b} \rightarrow b\bar{b}\ (t\bar t)$ and 
$q\bar{q} \rightarrow b\bar{b}\ (t\bar t)$ 
processes at the LHC for $M_V =1$ TeV as a function of $\Gamma_V$.}
\label{fig:bflhc}
\end{figure}
The higher energy of the LHC allows us to consider different types 
of signals in this case. We begin with the heavy-quark pair production 
processes at the Tevatron. We then discuss associated 
production of the new resonance with both  heavy quark pair 
$b\bar{b},\ t\bar t$ \cite{associated} 
and a single top \cite{singletop}. 
Although both of these processes have significantly 
smaller cross-sections than the Drell-Yan type of
heavy-quark pair signal, they also have 
much smaller backgrounds and a unique topology that 
offers a better chance for the signal observation.

\subsection{ $pp \rightarrow b\bar{b}X$ and $t\bar{t}X$}

The rates for the processes $pp \rightarrow b \bar{b}X,\  t\bar{t}X$ are much larger 
at the higher center of mass energies that can be reached at the LHC. It is also 
possible to search for heavier resonances for which the 
signal/background ratio is expected to be larger than what was 
possible at the Tevatron. We first explore the vector resonance. 
In Fig.~\ref{fig:bflhc}, 
we show the relative contributions to the signal from the 
$b\bar{b}$ annihilation process for $M_V =1$~TeV as a 
function  of $\Gamma_V$. We see that for both final states about 
$85\%$ of the signal events originate from $b\bar{b}$ annihilation 
if $\Gamma_V$ is  larger than $60$~GeV. In this case we are much less 
sensitive to the more model-dependent terms that arise from the light-quark 
annihilation processes. For illustration, we keep the same couplings 
that were used for the Tevatron studies in the last section: 
$g_V = 0.63$ for the vector and $\kappa_b = \kappa_t = 1$ for the 
scalar, as given in Eqs.~(\ref{cotr}) and (\ref{kappa}). 
The signal parameters we use for the LHC are thus,
\begin{equation}
M_R = 1~{\rm TeV},\ \Gamma_V =  127~{\rm GeV},\ \Gamma_S = 110~{\rm GeV}.
\label{lhcsp}
\end{equation}

\begin{figure}[t]
\begin{center}
\includegraphics[width = 14cm]{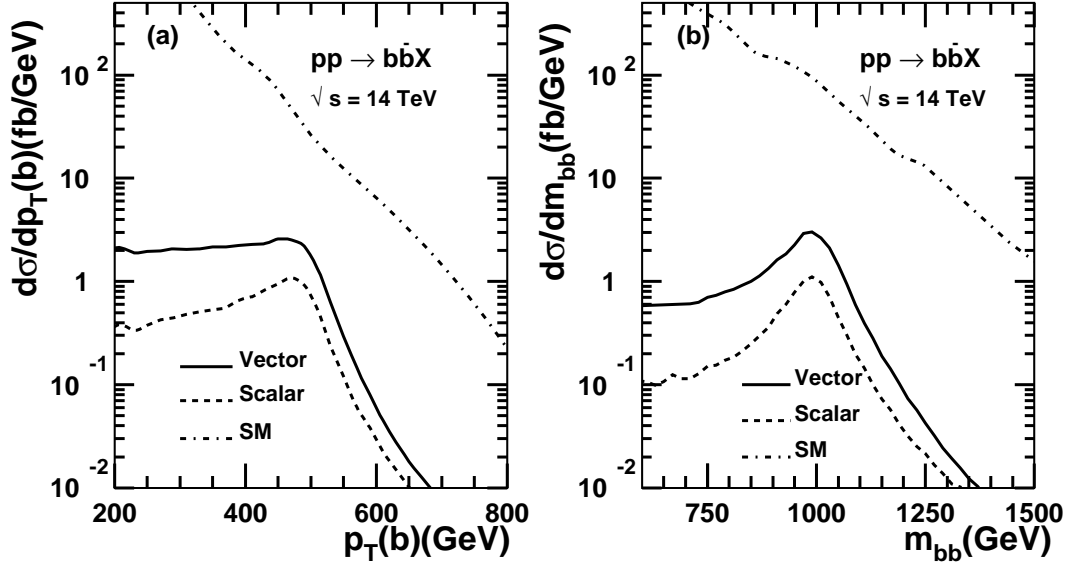}
\end{center}
\caption{(a) $p_T(b)$ and (b) $m^{}_{b\bar{b}}$ distributions for the 
signal and background at the LHC. The signal is calculated for 
the parameters in Eq.~(\ref{lhcsp})
and the dot-dashed curves indicate the standard model background. }
\label{fig:lhcpt-bb}
\end{figure}
\begin{figure}[t]
\begin{center}
\includegraphics[width = 14cm]{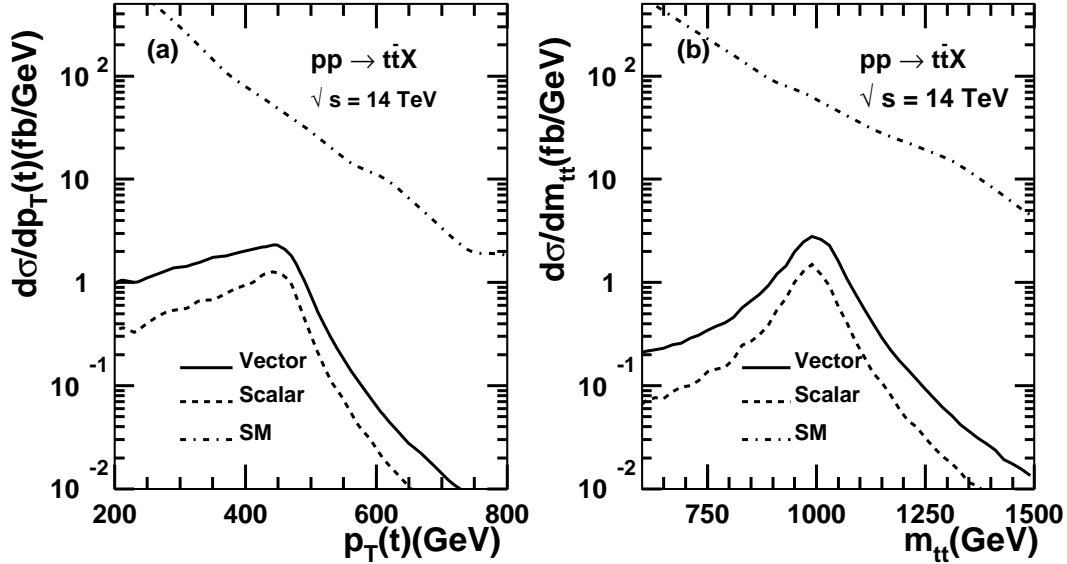}
\end{center}
\caption{(a) $p_T(t)$ and (b) $m^{}_{t\bar{t}}$ distributions for the 
signal and background at the LHC. For the signal we use the 
parameters of Eq.~(\ref{lhcsp}) and 
the dot-dashed curves indicate the standard model background. }
\label{fig:lhcpt}
\end{figure}
\begin{figure}[t]
\begin{center}
\includegraphics[width=16cm]{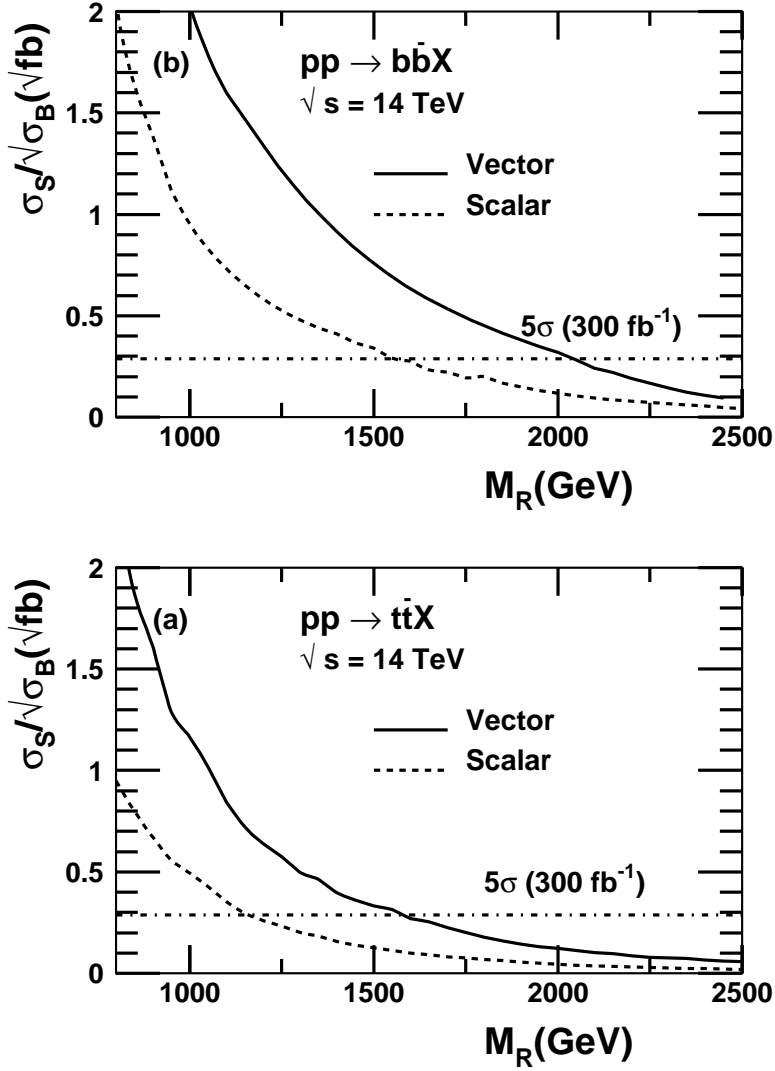}
\end{center}
\caption{ 
Statistical sensitivity at the LHC with  an integrated luminosity of 300~fb$^{-1}$ 
to a new resonance as a function of the 
resonance mass for (a) $b\bar b$ production, and (b) $t\bar t$ production. 
The solid curve is for a vector resonance and the dotted curve for a scalar 
resonance. The dot-dashed line indicates the $5\sigma$ sensitivity level. 
The couplings have been taken as in Eq.~(\ref{lhcsp}).}
\label{fig:lhcreach}
\end{figure}

For the processes $pp \rightarrow b\bar{b}X,\ t\bar tX$ at the LHC
we evaluate both the signal and background with the basic cuts 
\begin{eqnarray}
p_T(b,\ t) > 200 {\rm ~GeV},  \quad  |y_{b, t}| & <  2,
\label{lhc-cuts}
\end{eqnarray}
as well as with the cuts of Eq.~(\ref{tev-cuts1}) to optimize 
the signal/background ratio. In Fig.~\ref{fig:lhcpt-bb} 
we show the $p_T(b)$ and $m_{b\bar{b}}$ distributions for $pp \rightarrow b\bar{b}X$. 
The corresponding distributions for $pp \rightarrow t\bar{t}X$ are presented 
in Fig.~\ref{fig:lhcpt}. 
We see from these figures that the background at the LHC is about 
one order of magnitude larger than the signal. This is already better 
than the situation for the lower resonance mass at the Tevatron. 
The statistical sensitivity for the signals at the LHC is summarized 
in Fig.~\ref{fig:lhcreach}. 
We have used the same tagging rates and efficiencies as in the Tevatron. 
The dot-dashed lines indicate the $5\sigma$ sensitivity
assuming a total integrated luminosity of 300~fb$^{-1}$. 
The figure shows that the sensitivity extends 
to about $M_R\approx 2$ TeV at a $5\sigma$ level, for both vector and
scalar signals and for both $b\bar b$ and $t\bar t$ channels. 
Although the situation looks much more promising than at the Tevatron, 
due to the much larger luminosity and production cross-section, 
we must still bear in mind that the signal/background ratio is 
small and that more realistic simulations including detector effects are necessary
to reach definitive conclusions.

\subsection{$pp \to b\bar{b}t\bar{t}X$ }
\begin{figure}[t]
\begin{center}
\includegraphics[width=16cm]{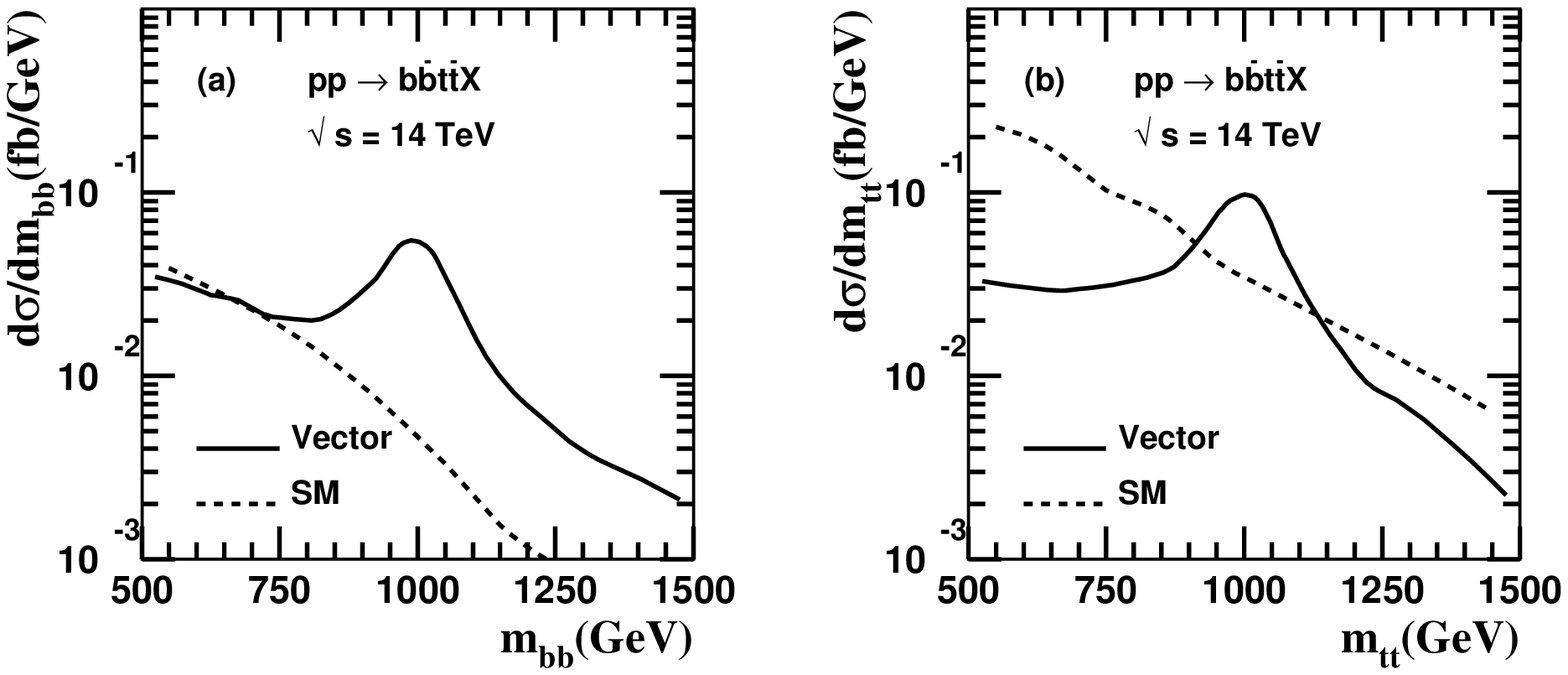}
\end{center}
\caption{(a) $m_{b\bar{b}}$ and (b) $m_{t\bar{t}}$
distributions for a vector resonance with $M_V = 1$~TeV and  
$\Gamma_V = 127$~GeV
(solid curves)  and the SM background (dashed curves)
in $pp \to b\bar{b}t\bar{t}X$ at the LHC.} 
\label{fig:bbtt-v}
\end{figure}

\begin{figure}[t]
\begin{center}
\includegraphics[width=14cm]{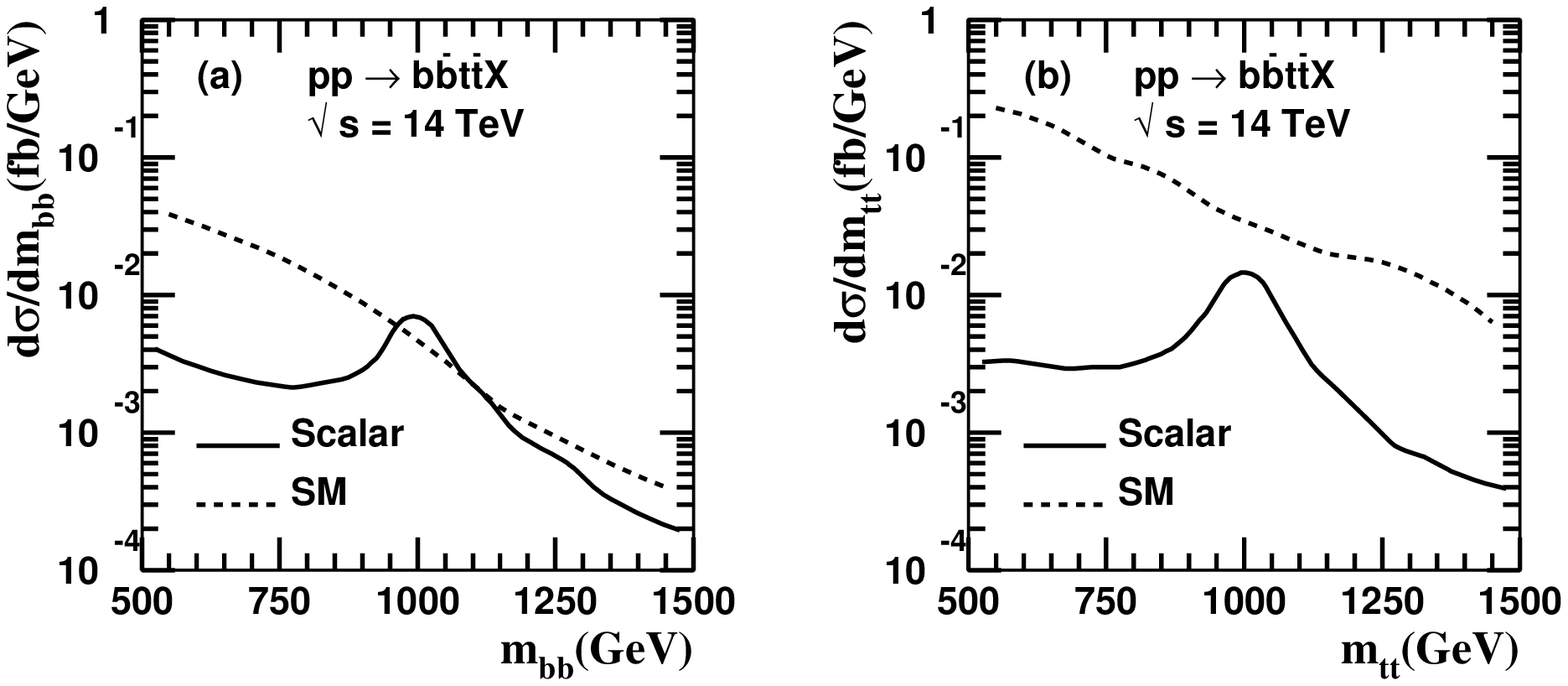}
\end{center}
\caption{(a) $m_{b\bar{b}}$ and (b) $m_{t\bar{t}}$
distributions for a scalar resonance with $M_S = 1$~TeV and  
$\Gamma_S = 110$~GeV
(solid curves) and the SM background (dashed curves)  in 
$p p \to b\bar{b}t\bar{t}X$ at the LHC.} 
\label{fig:bbtt-s}
\end{figure}

The much larger phase space available at the LHC permits us 
to explore more complicated processes. In particular processes 
with four heavy quarks, originating in the production 
of the heavy resonance in association with two heavy quarks, 
have been found to be very useful in Higgs studies \cite{associated}. 
At LHC energies these processes would be dominated by 
the initial subprocess $gg\to b\bar b$ followed by a heavy resonance 
radiated off one of the $b$ quarks. Since this coupling is enhanced 
in the new physics scenario that we are considering, this process 
could be significantly large. In addition, its unique topology 
could make the large QCD backgrounds more manageable. In this 
section we investigate this possibility. We use the program 
COMPHEP \cite{comphep} to compute the signal cross-sections for this process. 

In the process $pp \rightarrow b\bar{b} t \bar{t} X$ one of the $b$ quarks 
radiates a heavy resonance which decays to $t \bar t$. There is also 
a contribution from $gg\to t \bar t R$ followed by a decay $R \to b \bar b$. 
The signal is completely dominated by the gluon fusion process. There is also a  
much smaller contribution initiated by $b\bar b$ annihilation that we have calculated 
but not included. 
For illustration we use the same model-parameters as in the previous 
section, Eq.~(\ref{lhcsp}). We implement basic cuts 
\begin{eqnarray}
p_T(b) > 100 {\rm ~GeV}, \quad p_T(t) > 50 {\rm ~GeV}, \quad  |y_b|  <  2.
\label{bbtt-cuts}
\end{eqnarray}
\begin{figure}[tb]
\begin{center}
\includegraphics[width=16cm]{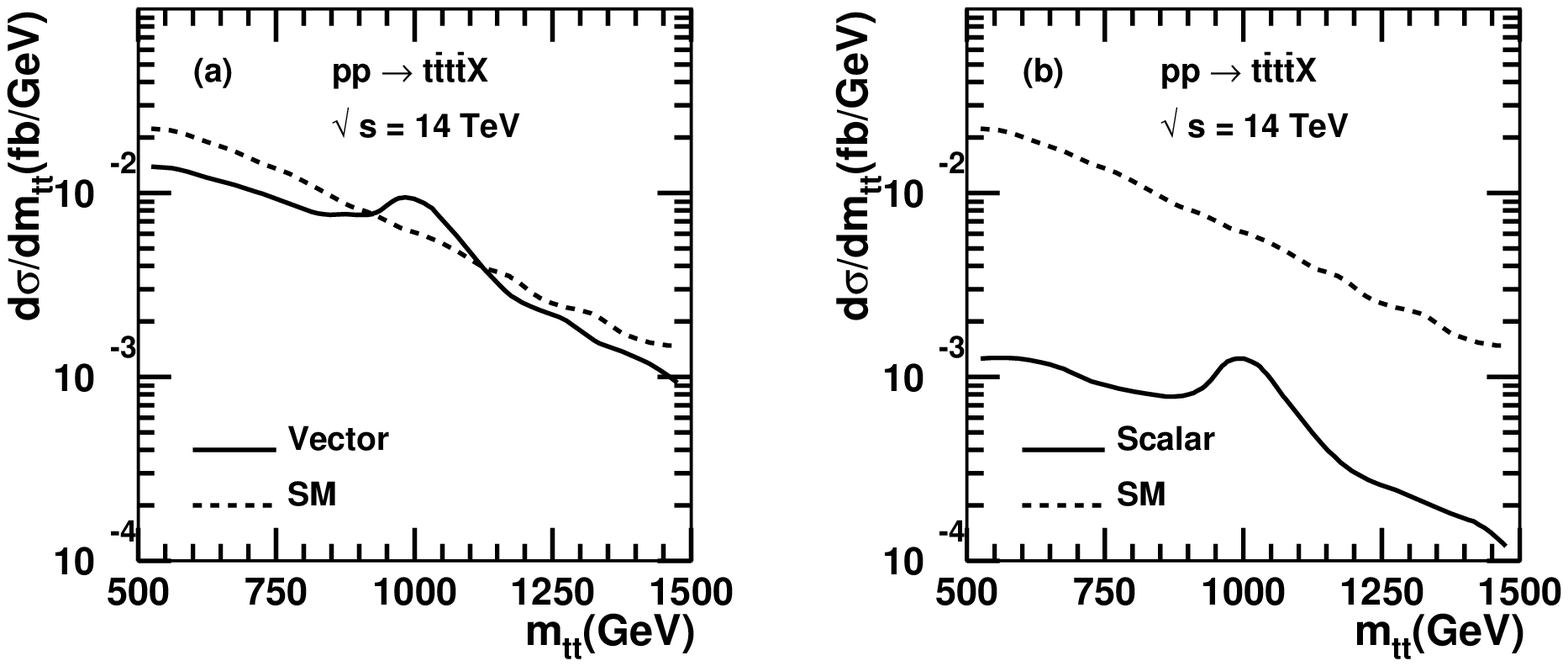}
\end{center}
\caption{
$m_{t{t}}$ distributions for $p p \to t\bar{t}t\bar{t}X$ at the LHC for 
(a) a vector resonance and (b) a scalar resonance with the parameters 
of Eq.~(\ref{lhcsp}). The SM background is depicted by the dashed curve.}
\label{fig:4t}
\end{figure}
In Figs.~\ref{fig:bbtt-v} and \ref{fig:bbtt-s}, we show the  $m_{b\bar{b}}$ 
and $m_{t\bar{t}}$ distributions for the process $p p \to b\bar{b}t\bar{t}X$ at the LHC 
for a vector and a scalar resonances respectively.   
As expected, there are peaks in these distributions originating in the 
resonance. 
It is particularly encouraging to see that the signal peaks are above the 
continuum background making a 
signal observation more promising than in the channels studied 
in the previous sections. 
With the parameters we have chosen, the vector resonance has a larger 
production rate than the scalar resonance and therefore results in a larger 
reach for the LHC. This however, mostly reflects the fact that we have chosen 
somewhat weaker couplings for the scalar. An additional cut 
is used to optimize the signal/background ratio,
\begin{eqnarray}
M_R - 4\Gamma_{R}  <  m_{b\bar{b}, t\bar{t}} <  M_R + 4\Gamma_{R}.
\label{lhc-cuts3}
\end{eqnarray}
 
\begin{figure}[tb]
\begin{center}
\includegraphics[width=16cm]{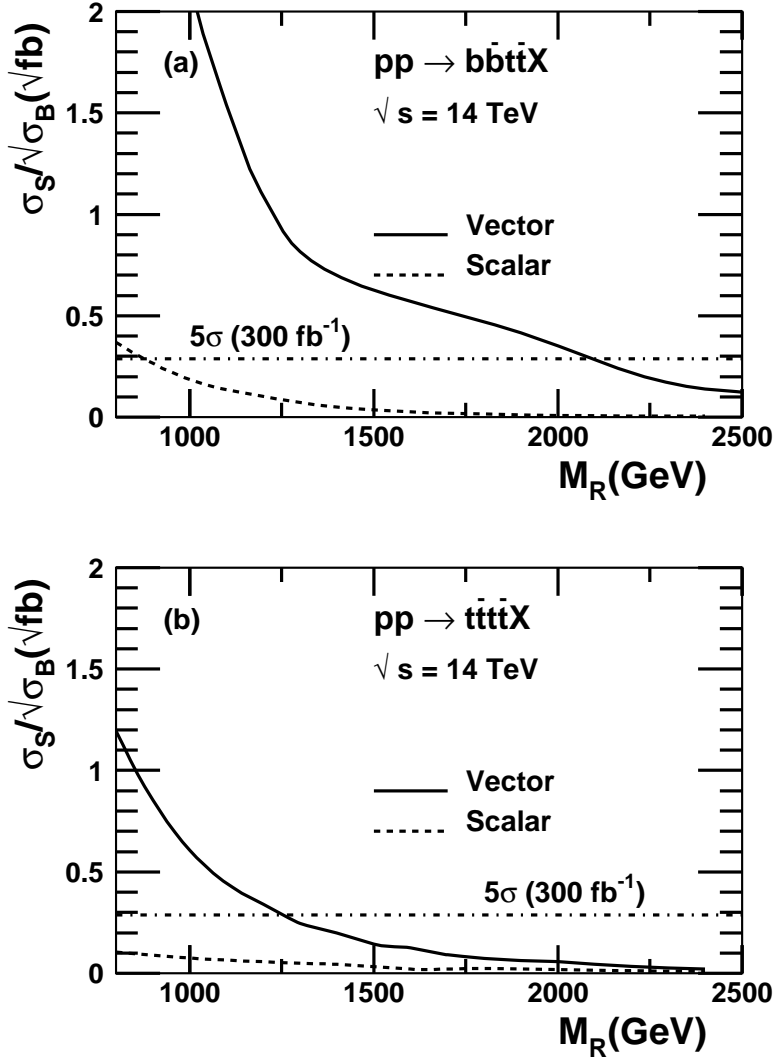}
\end{center}
\caption{ 
Statistical sensitivity at the LHC with  an integrated luminosity of 300 fb$^{-1}$ 
to a new resonance as a function of the 
resonance mass for (a) $pp \rightarrow b \bar{b} t\bar{t}X$,  
and (b) $pp \rightarrow t \bar{t} t\bar{t}X$. 
We use the parameters given in Eq.~(\ref{lhcsp}).} 
\label{fig:reach-4t-2t2b}
\end{figure}

Applying this cut to both $m_{b\bar{b}}$ and $m_{t\bar{t}}$, 
we present the sensitivity at the LHC assuming a total integrated luminosity of 
300 fb$^{-1}$ in Fig.~\ref{fig:reach-4t-2t2b}. We have used the same 
identification efficiencies for the $b$ and $t$ quarks as before. We 
see that a $5\sigma$ sensitivity may be reached for masses up to 2~TeV 
for the vector resonance. 

\subsection{$pp \to t\bar{t}t\bar{t}X$ }

The heavy resonance can also be produced in association with a $t\bar{t}$ 
pair. The dominant production mechanism for this mode is again gluon fusion.  
We have also calculated the $b\bar{b}$ annihilation mechanism but found it to be two 
orders of magnitude smaller and do not include it. We compute our signals with the 
program COMPHEP \cite{comphep}. 
Fig.~\ref{fig:4t} shows the $m_{t\bar{t}}$ distributions for 
new resonances with the same parameters as in Eq.~(\ref{lhcsp}).
For the new scalar 
this channel is sensitive to $\kappa_t^2$ whereas the 
$pp \to b\bar{b}t\bar{t}X$ channel is sensitive to $\kappa_t \kappa_b$. This 
distinction is important in models where only one of these couplings is 
large. We use the same basic cuts as in  
Eq.~(\ref{bbtt-cuts}). To construct the $m_{t\bar{t}}$ variable we 
have selected one $t$ and one $\bar{t}$ randomly. We therefore assume that 
it is possible to distinguish the $t$ from the $\bar{t}$ via their
leptonic decays.  
Using Eq.~(\ref{lhc-cuts3}), we present the statistical sensitivity 
of this process in Fig.~\ref{fig:reach-4t-2t2b} using 
16\% event efficiency for {\it each} $t\bar{t}$ pair. 
Due to the lower cross-section and identification efficiency 
the reach in this channel is somewhat smaller than in the  
$p{p} \to b\bar{b}t\bar{t}X$ channel. 

\subsection{$Wb \to t, b\bar{b};\ t, t\bar{t}$ }

\begin{figure}[t]
\begin{center}
\includegraphics[width=14cm]{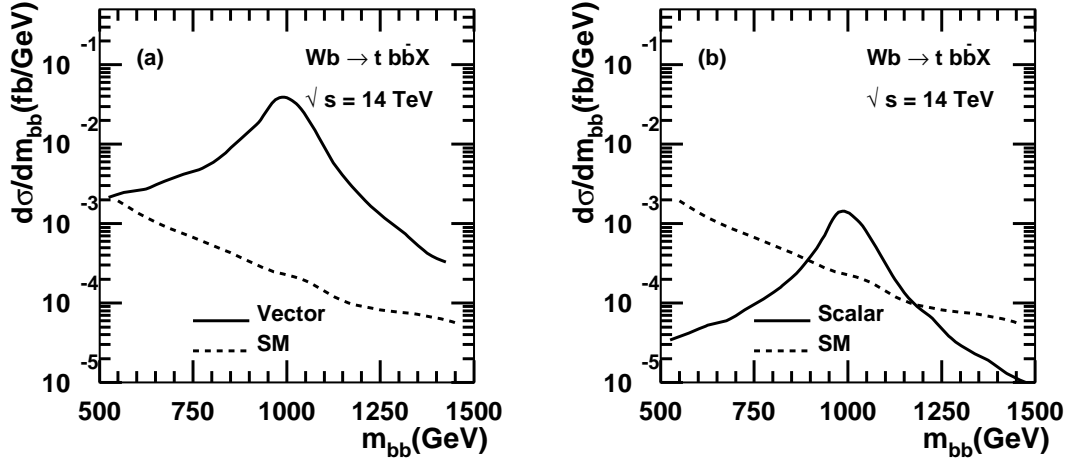}
\end{center}
\caption{ $m_{b{b}}$ distributions in $Wb \to t b\bar{b}$ at the LHC 
for (a) a vector resonance and for (b) a scalar resonance with 
the parameters of Eq.~(\ref{lhcsp}). 
The SM background is indicated by the dashed curve.}
\label{fig:wtbb}
\end{figure}

It is well known that single top quark production via the 
electroweak process $Wb \rightarrow t$ can be sizable 
due to the enhanced longitudinal gauge boson coupling
$W_Ltb$ at high energies \cite{singletop}. 
The cross section for single top production increases with energy 
up to about one-third of the cross section for 
$t\bar{t}$ pair production \cite{sintop}. The main 
advantage of this channel is the substantially smaller standard model 
background.
We now consider the effect of our new resonances on this process using 
the program COMPHEP \cite{comphep} to compute the signals. For this case 
only, we also use COMPHEP to estimate the standard model background.

We first consider the $Wb \to t b\bar{b}$ process with the basic cuts 
\begin{eqnarray}
p_T(t, b) > 100 {\rm ~GeV},  \quad |y_{t,b}| & < 2.
\label{wtbb-cuts}
\end{eqnarray}
The high $p_T$ cut is imposed  on {\it all} heavy quarks,
including the two b quarks that reconstruct the resonance mass as well as the single top quark. 

In Fig.~\ref{fig:wtbb}, we show the $m_{b\bar{b}}$ distribution. We see 
that the signals can be significantly above the SM background in this case. 
We use the cuts of Eq.~(\ref{lhc-cuts3}) to estimate the sensitivity 
to the new physics. 

\begin{figure}[tb]
\begin{center}
\includegraphics[width=14cm]{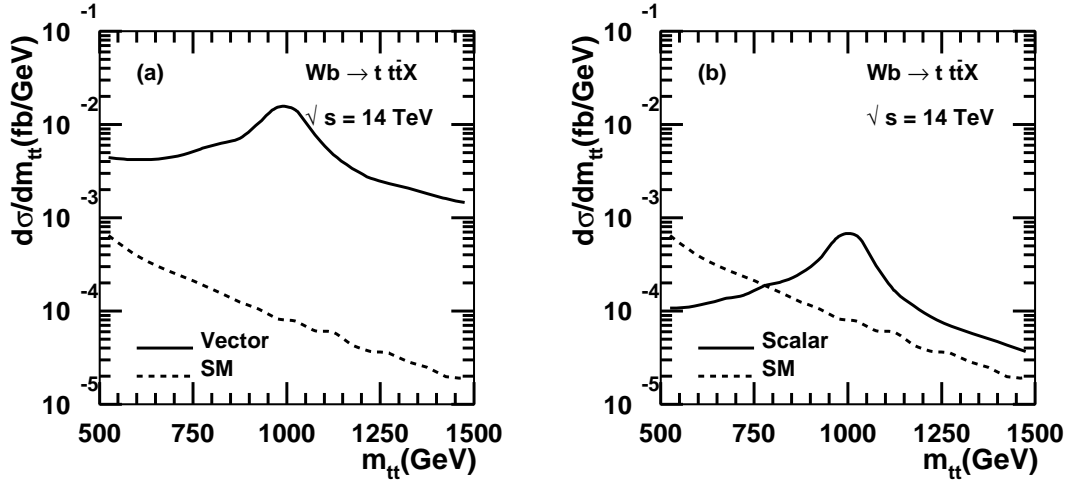}
\end{center}
\caption{
$m_{t{t}}$ distribution in the process $Wb \to tt\bar{t}$ at the LHC 
for (a) a vector resonance and for (b) a scalar resonance with 
parameters given in Eq.~(\ref{lhcsp}). The SM background is indicated by the 
dashed line.} 
\label{fig:wttt}
\end{figure}

\begin{figure}[t]
\begin{center}
\includegraphics[width=16cm]{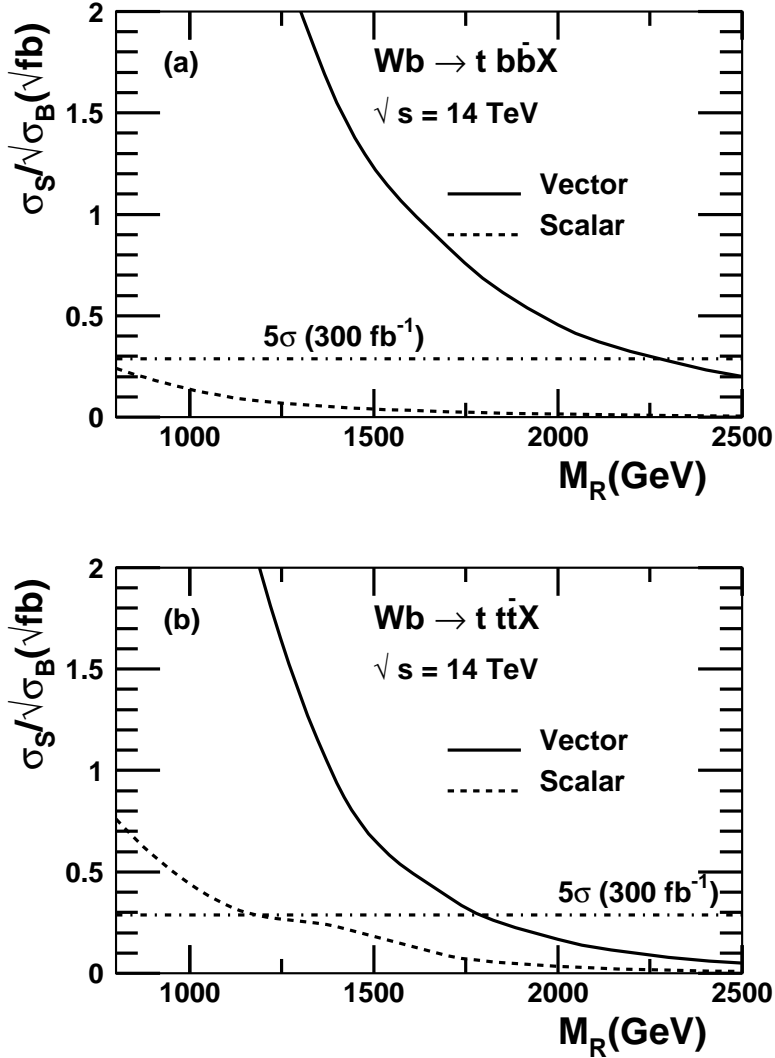}
\end{center}
\caption{
Statistical sensitivity at the LHC with  an integrated luminosity of 300 fb$^{-1}$ 
to a new resonance as a function of the 
resonance mass through the processes 
(a) $Wb \to t b\bar{b}$, 
and (b) $Wb \to t t\bar{t}$.
The solid curve is for a vector resonance and the dotted curve for a scalar 
resonance with parameters as in Eq.~(\ref{lhcsp}). 
The dot-dashed line indicates the $5\sigma$ sensitivity level.}
\label{fig:reach-wtb}
\end{figure}

For the process $Wb \to t t\bar{t}$ the standard model background is 
even smaller. The basic cut $|y_t| < 2$ leads us to the results in 
Fig.~\ref{fig:wttt}.  This process has the largest signal/background ratio 
of all the ones we have considered.   
Using the same $b$ and $t$ efficiencies that we used for the Tevatron we show 
in Fig.~\ref{fig:reach-wtb} the reach in this case. 

\section{Conclusion}

We have presented a preliminary study of signals for new resonances 
coupling to heavy quarks at the Tevatron and the LHC. We have considered 
a scalar and a vector resonance with masses of 400~GeV for the 
Tevatron study and 1~TeV for the LHC study and we have chosen  
couplings so that the resonances couple strongly to the top quark 
but are sufficiently narrow to be described by a Breit-Wigner shape. 
These parameters satisfy the existing constraints.

The most direct production mechanism is the 
Drell-Yan  process for  $pp\to b\bar b X$ and $ t\bar t X$.
At the Tevatron, due to the rather low production rate and substantial 
background from the heavy quark production, only a weak $3\sigma$ 
bound may be put on the vector state signal for $M_V<400$ GeV. 

At the LHC, on the other hand, the situation can be significantly improved. 
A $5\sigma$ statistical sensitivity may be reached for both a vector 
and a scalar state with $M_R\sim 1.5$ TeV. However, the large QCD backgrounds 
for the heavy quark pair production still 
lead to a low signal-to-background ratio (less than 10$\%$) near the 
resonance peaks. This renders the signal observation systematically
difficult. 

We found that  the most promising channels 
for the signal searches are multiple heavy quark production. The first
class of processes is essentially due to the $gg\to b\bar b,\ t\bar t$ 
with a heavy 
resonance radiation off a heavy quark leg. Consequently, the 4-quark
signals $b\bar b t\bar t$ and $t\bar t t\bar t$ would have much less
severe SM backgrounds. It is very interesting to note that the electroweak
process  $Wb\to t$, again associated with a heavy resonance radiation
off the top quark, could lead to very strong signal as well.
The reach can be up to $M_V \sim 2$~TeV with a $5\sigma$ significance.

More detailed studies including detector issues would be needed to reach 
more definitive conclusions and to fully determine the range of parameters 
that can be probed by the LHC.

\vskip 0.3cm

{\it Acknowledgments}: The work of T.H. was supported in part 
by the US DOE under contract No. DE-FG02-95ER40896, in part by the
Wisconsin Alumni Research Foundation, and in part by
National Natural Science Foundation of China.
The work of G.V. and Y. W. was supported
in part by DOE under contact number DE-FG02-01ER41155. 



\begin{thebibliography}{99}
 
\bibitem{chano}
M.~S.~Chanowitz,
Phys.\ Rev.\ Lett.\  {\bf 87}, 231802 (2001); 
M.~S.~Chanowitz,
Phys.\ Rev.\ D {\bf 66}, 073002 (2002)
[arXiv:hep-ph/0207123].

\bibitem{topcolor}
C.~T.~Hill,
Phys. Lett. {\bf B266}, 419 (1991);
{\it ibid.} {\bf B345}, 483 (1995);
E.~Eichten and K.~Lane,
Phys.\ Lett.\ {\bf B352}, 382 (1995).

\bibitem{han}
T.~Han, Y.~J.~Kim, A.~Likhoded and G.~Valencia,
Nucl.\ Phys.\ {\bf B593}, 415 (2001); 
T.~Han, D.~Rainwater and G.~Valencia,
Phys.\ Rev.\ D {\bf 68}, 015003 (2003)
[arXiv:hep-ph/0301039].


\bibitem{zprime}
X.~G.~He and G.~Valencia,
Phys.\ Rev.\ D {\bf 66}, 013004 (2002)
[Erratum-ibid.\ D {\bf 66}, 079901 (2002)]; 
X.~G.~He and G.~Valencia,
Phys.\ Rev.\ D {\bf 68}, 033011 (2003)
[arXiv:hep-ph/0304215];
X.~G.~He and G.~Valencia,
arXiv:hep-ph/0404229.


\bibitem{topac}
S.~Dawson and G.~Valencia,
Nucl.\ Phys.\ B {\bf 348}, 23 (1991);
R.~D.~Peccei, S.~Peris and X.~Zhang,
Nucl.\ Phys.\ B {\bf 349}, 305 (1991);
S.~Dawson and G.~Valencia,
Phys.\ Rev.\ D {\bf 53}, 1721 (1996)
[arXiv:hep-ph/9510455];
F.~Larios and C.~P.~Yuan,
Phys.\ Rev.\ D {\bf 55}, 7218 (1997)
[arXiv:hep-ph/9606397];
F.~Larios, M.~A.~Perez and C.~P.~Yuan,
Phys.\ Lett.\ B {\bf 457}, 334 (1999)
[arXiv:hep-ph/9903394].

\bibitem{hillparke}
C.~T.~Hill and S.~J.~Parke,
Phys.\ Rev.\ D {\bf 49}, 4454 (1994).

\bibitem{bess}
R.~Casalbuoni {\it et al.},
Phys.\ Lett.\ {\bf B155}, 95 (1985);
Nucl.\ Phys.\ {\bf B282}, 235 (1987);
{\bf B310}, 181 (1988);
Phys.\ Lett.\ {\bf B249}, 130 (1990);
{\bf B253}, 275 (1991).


\bibitem{unit}
B.~W.~Lee, C.~Quigg, and H.~B.~Thacker,
Phys.\ Rev.\ {\bf D16}, 1519 (1977);
M.~S.~Chanowitz, M.~A.~Furman and I.~Hinchliffe,
Nucl.\ Phys.\ B {\bf 153}, 402 (1979);
T.~Appelquist and M.~S.~Chanowitz,
Phys.\ Rev.\ Lett.\  {\bf 59}, 2405 (1987)
[Erratum-ibid.\  {\bf 60}, 1589 (1988)];
W.~J.~Marciano, G.~Valencia and S.~Willenbrock,
Phys.\ Rev.\ D {\bf 40}, 1725 (1989);
S.~Jager and S.~Willenbrock,
Phys.\ Lett.\ B {\bf 435}, 139 (1998)
[arXiv:hep-ph/9806286].

\bibitem{rate}  Elizaveta Chabalina, private communication.

\bibitem{Stelzer:1994ta}
T.~Stelzer and W.~F.~Long,
Comput.\ Phys.\ Commun.\  {\bf 81}, 357 (1994)
[arXiv:hep-ph/9401258].


\bibitem{associated}
Z.~Kunszt,
Nucl.\ Phys.\ B {\bf 247}, 339 (1984);
W.~J.~Marciano and F.~E.~Paige,
Phys.\ Rev.\ Lett.\  {\bf 66}, 2433 (1991);
J.~F.~Gunion,
Phys.\ Lett.\ B {\bf 261}, 510 (1991).

\bibitem{singletop}
S.~S.~D.~Willenbrock and D.~A.~Dicus,
Phys.\ Rev.\ D {\bf 34}, 155 (1986);
C.~P.~Yuan,
Phys.\ Rev.\ D {\bf 41}, 42 (1990);
S.~Cortese and R.~Petronzio,
Phys.\ Lett.\ B {\bf 253}, 494 (1991);
R.~K.~Ellis and S.~Parke,
Phys.\ Rev.\ D {\bf 46}, 3785 (1992);
F.~Maltoni, K.~Paul, T.~Stelzer and S.~Willenbrock,
Phys.\ Rev.\ D {\bf 64}, 094023 (2001)
[arXiv:hep-ph/0106293];
T.~Tait and C.~P.~Yuan,
Phys.\ Rev.\ D {\bf 63}, 014018 (2001)
[arXiv:hep-ph/0007298];
E.~Boos, L.~Dudko and T.~Ohl,
Eur.\ Phys.\ J.\ C {\bf 11}, 473 (1999)
[arXiv:hep-ph/9903215].

\bibitem{comphep}
A.~Pukhov {\it et al.},
arXiv:hep-ph/9908288.


\bibitem{sintop} 
T. Stelzer, Z. Sullivan and S. Willenbrock, 
Phys.\ Rev.\ {\bf D58}, 94021(1998); 
M. Beneke {\it et al}., in ``Proceedings of the Workshop 
on Standard Model Physics (and more) at the LHC''; 
M.~Beneke {\it et al.},
arXiv:hep-ph/0003033.


\end{thebibliography}
\end{document}